\documentclass[journal=jpcl,manuscript=article]{achemso}

\usepackage{chemformula} % Formula subscripts using \ch{}
\usepackage[T1]{fontenc} % Use modern font encodings

\author{Zikun Zhuang}
\author{Chengdong Yang}
\affiliation{Department of Chemistry, College of Science, Southern University of Science and Technology, Shenzhen, 518055, China}
\author{Yuchen Wang}
\affiliation{Department of Chemistry and The James Franck Institute, The University of Chicago, Chicago, Illinois 60637, USA}
\author{Dong H. Zhang}
\affiliation{State Key Laboratory of Chemical Reaction Dynamics, Dalian Institute of Chemical Physics, Chinese Academy of Sciences, Dalian, 116023, China}
\alsoaffiliation{Hefei National Laboratory, Hefei, 230088, China}
\author{Bin Zhao}
\email{zhaobin@sustech.edu.cn}
\affiliation{Department of Chemistry, College of Science, Southern University of Science and Technology, Shenzhen, 518055, China}

\title{Quantum Mechanical Studies of Photodissociation Dynamics on Quantum Computers}
\abbreviations{IR,NMR,UV}
\usepackage{graphicx}
\usepackage{braket}
\usepackage{amsmath}
\usepackage{float}
\usepackage{subfigure}
\usepackage[normalem]{ulem}
\usepackage{xcolor}
\usepackage{soul}
\usepackage{tikz}
\usetikzlibrary{quantikz}

\begin{document}

\begin{tocentry}
  \centering
  \includegraphics[width=\textwidth]{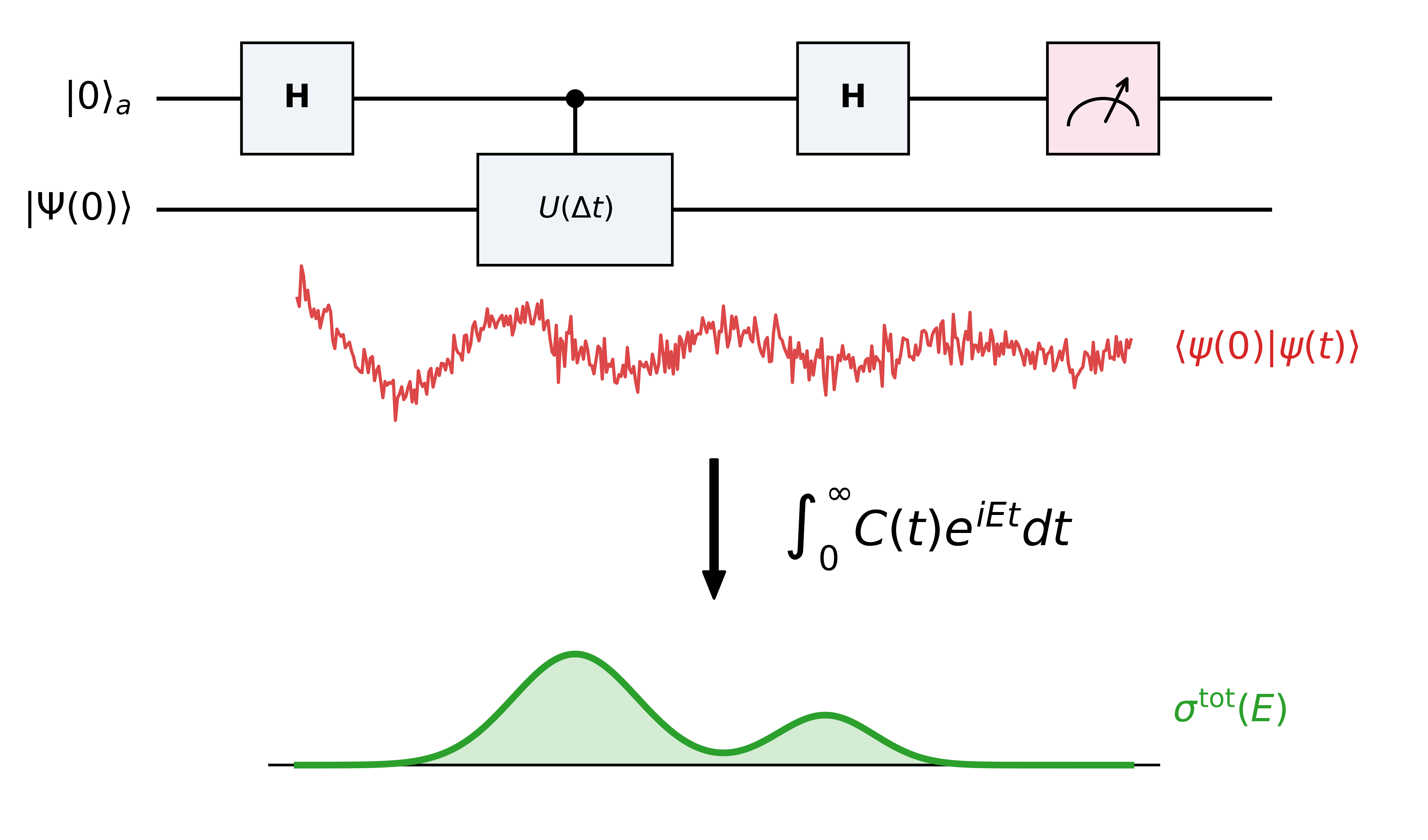}
\end{tocentry}


\newpage
\begin{abstract}
Theoretical quantum dynamics calculations scale deeply with system size, rendering classical calculations intractable for complex systems.
While quantum computing offers a natural solution, its application to nuclear quantum dynamics remains scarce.
Here, we present a quantum algorithm to study  
photodissociation dynamics on quantum computers, benchmarked on the NOCl molecule.
The wavefunction is propagated via a split-operator method, utilizing the Quantum Fourier Transform and unitary transformation matrix to switch representations.
To impose outgoing boundary conditions on a truncated grid, we use a non-unitary absorbing potential propagator, implemented through a dilation scheme.
The photodissociation cross section is calculated from the auto-correlation function, which is extracted using the Hadamard test.
Our quantum computing results agree well with benchmarks under ideal conditions, and we further demonstrate that the algorithm is robust to noise and statistical sampling errors, indicating the promising application of noisy devices to quantum dynamics studies.

\end{abstract}
\newpage
\section{Introduction}

Quantum reaction dynamics is fundamentally important for understanding the essential intricacies of chemical reactions at the most detailed level.\cite{schatz_spiers_2024, pan_state--state_2023, pan_crossed_2017} 
Quantum dynamics studies are the theoretical framework for elucidating
molecular-level chemical transformations by treating nuclear motion quantum-mechanically,\cite{zhang_recent_2016, li_advances_2020}
revealing quantum features that exceed the predictive limitations of classical and semi-classical methods.
For example, quantum effects such as 
quantum tunneling\cite{yang_enhanced_2019}, 
geometric phase effects\cite{yuan_observation_2018}, 
stereodynamics\cite{wang_stereodynamical_2023,yoder_steric_2010, wang_steric_2011}
quantum interference\cite{berry1984quantal, xie_quantum_2020,zhou_quantum_2021}, 
dynamic resonances\cite{qiu_observation_2006,wang_dynamical_2013,yang_extremely_2015}
and non-adiabatic transitions \cite{curchod_ab_2018, guan2021high}
all play crucial roles in determining the mechanistic reaction pathways, reaction rates, and product quantum state and angular distributions.
These effects can only be faithfully captured by full-quantum molecular dynamics calculations on highly accurate potential energy surfaces (PESs). 

Quantum molecular dynamics calculations are among the many challenges
in computational chemistry. 
Understanding these dynamics at the quantum-mechanical level requires solving the time-dependent Schrödinger equation for systems with many interacting particles, 
a task that becomes intractable on classical computers.
The core challenge arises from the exponential scaling of computational resources required to represent and evolve quantum many-body wavefunctions. 
For a system with $d$ degrees of freedom, the wavefunction is represented by $N^d$ coefficients, where $N$ is the number of basis functions per degree of freedom. 
This exponential scaling, often called the “curse of dimensionality”, imposes severe restrictions on the size of systems that can be treated accurately. 
Consequently, exact quantum reaction dynamics calculations are typically limited to systems with about 5-6 atoms,\cite{tan_revealing_2025,chen_reactivity_2020}
obviously excluding most realistic reactions in chemistry and biology. 
While more efficient methods like the Multi-Configuration Time-Dependent Hartree (MCTDH)\cite{meyer_multi-configurational_1990} approach have extended these limits, 
it still faces fundamental challenges to large systems \cite{beck_multiconfiguration_2000,manthe_multilayer_2008,shi_full_2025,shi_quantum_2023}

Quantum computing has the potential to surmount the exponential scaling limitations that fundamentally constrain classical computational frameworks. 
The inherent correspondence between quantum hardware and quantum dynamics naturally supports accurate modeling of quantum coherence and exponential scaling of Hilbert space beyond the reach of classical computers. 
Quantum chemistry is one of the most promising applications of quantum computing.\cite{cao_quantum_2019,schleich_cracking_2025, alexeev2025perspective}. 
By mapping the molecular wavefunction onto $m$ qubits with a Hilbert space size of $2^m$, quantum algorithms enable the simulation of complex systems with memory and time requirements that scale polynomially with system 
size\cite{zalka_simulating_1998,kassal_polynomial-time_2008}. 
Current applications in chemistry have focused largely on electronic 
structure problems. \cite{aspuru-guzik_simulated_2005,wang_quantum_2008,peruzzo_variational_2014}. 

In contrast, application to reaction dynamics remains relatively unexplored. 
This disparity arises from the complexity of calculating coupled electronic-nuclear motion.
Thanks to the Born-Oppenheimer approximation, which decouples electronic and nuclear motions, quantum dynamics studies can typically be divided into constructing PESs and propagating nuclear wavefunctions thereafter. 
For example, photodissociation dynamics \cite{von_rague_schleyer_photodissociation_1998} involves photon absorption that promotes the molecule from the ground electronic state to the dissociative excited electronic state, necessitating the construction of multiple PESs and the subsequent time-dependent propagation of nuclear wavefunctions. 

Real-time wavefunction propagation is a natural candidate for quantum computation 
\cite{zalka_simulating_1998}, 
but translating a propagator into a quantum 
circuit requires substantial circuit depths and gate counts, which often exceed the 
coherence time limits and the fidelity of near-term hardware 
\cite{schleich_cracking_2025}.
One of the challenges originates from encoding the high-dimensional potential energy operator into practical quantum circuit architectures.
Consequently, implementing high-fidelity, 
long-time dynamics for realistic systems remains a challenge 
on noisy quantum hardware. 
Nonetheless, notable recent studies have
established the feasibility of such calculations in the quantum computing 
framework, 
such as the efficient evaluation of $S$-matrices by Xing et al. \cite{xing_hybrid_2023}, 
the calculation of reaction probabilities via M{\o}ller operators by Kale and Kais \cite{kale_simulation_2024}, and the simulation of absorption spectra by Wan et al. \cite{wan_hybrid_2024} and Feng et al. \cite{feng_quantum_2025}.
These recent studies collectively demonstrate the potential application of quantum computation to quantum dynamics, but they primarily focus on systems governed by unitary operations. 

However, in the traditional time-dependent wavepacket (TDWP) 
method, not all operators are unitary. For example, calculations of dissociative processes are performed within a finite coordinate range and the complex absorbing potential (CAP) can be used to enforce the outgoing boundary condition at the coordinate edge, otherwise the dissociating wavefunction would reach the grid boundaries and introduce unphysical artifacts that  
corrupt the extraction of subsequent dynamics information. 
The operation of the absorbing potential in the time propagator is non-unitary, which poses a practical challenge to quantum computation. 
One possible approach is to use a sufficiently large coordinate range so that the wavefunction propagation completes before reaching the boundary,\cite{kale_simulation_2024} but this would increase the spatial discretization required to accurately represent and propagate the wavefunction.

In this work, we addressed the issue of the non-unitary propagator with a dilation strategy,\cite{hu2020quantum, schlimgen2021quantum, mangin-brinet_efficient_2024} 
using the photodissociation of the nitrosyl chloride (NOCl) molecule on its $S_1$ 
excited electronic state as a benchmark. 
The choice of NOCl as a target system is motivated by two factors. 
First, it serves as a well-established 
prototype for direct photodissociation, offering a clear physical picture 
of wavepacket motion on a repulsive surface. Second, NOCl has long been used as a 
standard benchmark system to validate the accuracy of theoretical methodologies 
\cite{manthe_wave-packet_1992,beck_multiconfiguration_2000}.
In the following sections, we implemented and verified the quantum computing algorithm through IBM's Qiskit 
\cite{qiskit2024} framework. We propagated the wavefunction on the $S_1$ state with 
CAP at the coordinate edge and calculated the auto-correlation function using the Hadamard test. In the results section, we compare the results from the quantum computing algorithm with the benchmark results, and check the robustness of the algorithm against the number of sampling shots and device noise. This study showcases the potential application of the current algorithm to extracting meaningful dynamical information from noisy quantum devices for realistic photodissociation processes.

\section{Theory}
The theory of TDWP method is well established for triatomic systems and its implementation on quantum computer follows a similar procedure. 
Here, the coordinate system, nuclear Hamiltonian, wavefunction representation, time propagation, absorbing potential, and cross sections calculation will be briefly introduced. 
Then, these traditional TDWP components are translated into a quantum computing algorithm, which inherently requires unitary operations.
In the following subsections, the theory of the traditional TDWP method is first described before the corresponding quantum computing algorithm is proposed.

\subsection{Coordinate system and Hamiltonian}

\begin{figure}[H]
    \centering
    \includegraphics[width=0.5\textwidth]{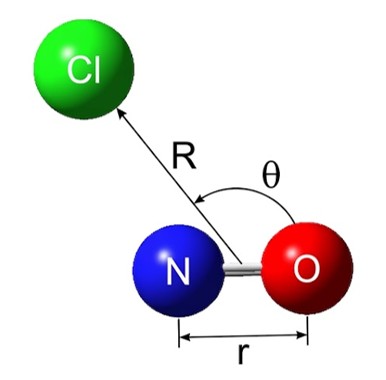}
    \caption{Jacobi coordinate system $(r, R, \theta)$ used for the 
    triatomic NOCl molecule. Here, $r$ represents the NO bond length, 
    $R$ is the distance from the Cl atom to the center of mass of the 
    NO fragment, and $\theta$ is the angle between the vectors $\mathbf{r}$ and $\mathbf{R}$.}
    \label{fig:jacobi}
\end{figure}
 
The nuclear Hamiltonian is given in the Jacobi coordinate system ($r$, $R$, $\theta$), defined in Fig. \ref{fig:jacobi}. Specifically, taking the NOCl molecule as an example, $r$ denotes the NO bond length, $R$ represents the distance from the Cl atom to the center of mass of the NO fragment, and $\theta$ is the enclosed angle between the vectors $\mathbf{r}$ and $\mathbf{R}$.
The Hamiltonian is then given as (atomic units are used unless otherwise specified):
\begin{eqnarray}
    \hat{H}_{0/1}&=&
    \hat{T}+\hat{V}_{0/1}(R,r,\theta) \nonumber \\
    &=&
    \hat{T}_R+\hat{T}_r+\hat{T}^R_{\mathrm{rot}}+\hat{T}^r_{\mathrm{rot}}+\hat{V}_{0/1}(R,r,\theta) \nonumber \\
    &=&
    -\frac{1}{2\mu_R}\frac{\partial^2}{\partial R^2}
    -\frac{1}{2\mu_r}\frac{\partial^2}{\partial r^2}
    +\frac{\hat{L}^2}{2\mu_R R^2}+\frac{\hat{j}^2}{2\mu_r r^2}+\hat{V}_{0/1}(R,r,\theta) 
  \label{eq:ham}
\end{eqnarray}
where the Hamiltonian $H_{0/1}$ correspond to the $S_0$ and $S_1$ electronic
states, respectively, with $\hat{V}_{\mathrm{0/1}}(R,r,\theta)$ as the corresponding potential energy operators.
$\mu_R$ and $\mu_r$ are the reduced masses 
associated with the dissociation coordinate $R$ and the vibrational 
coordinate $r$, respectively. 
The operators $\hat{L}$ and $\hat{j}$ represent the orbital 
and diatomic rotational angular momenta, respectively. 
If the total angular momentum is restricted to zero ($\hat{J}_{\mathrm{tot}}=0$), $\hat{L}=\hat{J}_{\mathrm{tot}}-\hat{j}=-\hat{j}$, 
yielding $\hat{L}^2 = \hat{j}^2$. This restriction simplifies the 
centrifugal terms in Eq. \ref{eq:ham} and will be used for the subsequent calculations.

The same coordinate system and Hamiltonian are used for both the traditional and quantum computing algorithm. 

\subsection{Wavefunction representation and time-evolution}
In the traditional TDWP method, the wavefunction $\psi(R,r,\theta)$ is usually represented using the discrete variable representation (DVR)
\cite{light_discrete-variable_2000}. 
The wavefunction in the radial coordinates $R$ 
and $r$ are discretized using a uniform grid derived from a Fourier basis, \cite{colbert_novel_1992}
while the one in the angular coordinate $\theta$ is discretized using a Legendre-DVR
basis. \cite{lill_discrete_1982}
The wavefunction $\psi(R,r,\theta)$ is thus expanded in terms of the basis functions in the following form:
\begin{equation}
    \psi(R,r,\theta) = \sum_{j_R} \sum_{j_r} \sum_{j_\theta}
    c_{j_R, j_r, j_\theta} \xi_{j_R}(R) \chi_{j_r}(r) \phi_{j_\theta}(\theta)
\end{equation}
where $\xi_{j_R}(R)$, $\chi_{j_r}(r)$, and $\phi_{j_\theta}(\theta)$ denote the basis
functions for each degree of freedom, and $c_{j_R, j_r, j_\theta}$ %ZZ's
are the expansion coefficients.

The Hamiltonian in Eq (\ref{eq:ham}) is also discretized in the above DVR basis functions. Importantly, the DVR matrix representations of the coordinate operators, such as the potential energy operator $\hat{V}$, the $1/R^2$ operator, and the $1/r^2$ operator, are diagonal.
The kinetic energy operators $\frac{\partial^2}{\partial R^2}$, $\frac{\partial^2}{\partial r^2}$, and $\hat{j}^2$ are not diagonal in DVR but are diagonal in the finite basis representation (FBR). \cite{light_generalized_1985}
The two representations can be transformed from one to the other by the discrete Fourier transform in the radial coordinates $R$ and $r$, and a unitary transformation matrix $U_{\theta}$
is employed to transform between the two representations in the angular coordinate $\theta$.

The real-time propagation of the wavefunction is implemented via the second-order
split-operator method \cite{feit_solution_1982,feit_solution_1983}.
For a short propagation time step $\Delta t$, the time evolution propagator
is approximated as:
\begin{equation}
    e^{-i\hat{H}\Delta t} \approx 
    e^{-i\hat{V}\Delta t/2}
    e^{-i\hat{T}\Delta t}
    e^{-i\hat{V}\Delta t/2}
  \label{eq:so}
\end{equation}
where $\hat{T}$ represents the kinetic energy operator and $\hat{V}$
is the potential energy operator.
Because $\hat{T}$ and $\hat{V}$ do not commute, the second-order 
split-operator approximation incurs a local truncation error of 
$O(\Delta t^3)$ \cite{feit_solution_1982}, commonly referred to as the Trotter error in the context of quantum simulation.

The quantum computing algorithms for the wavefunction representation and time evolution are straightforward.
On a digital quantum architecture, the multidimensional wavefunction is 
mapped onto discrete qubit registers following a standard binary encoding 
scheme\cite{zalka_simulating_1998}. 
Each degree of freedom ($i \in \{R, r, \theta\}$) is allocated to a specific quantum register comprising $n_i = \log_2 N_i$ qubits, where $N_i$ is the number of grid points for that coordinate.
The three-dimensional wavefunction is then mapped directly onto the 
computational basis of the joint quantum register as:
\begin{equation}
    |\Psi\rangle = \sum_{j_R=0}^{N_R-1} \sum_{j_r=0}^{N_r-1} 
    \sum_{j_\theta=0}^{N_\theta-1} c_{j_R, j_r, j_\theta} 
    |j_R\rangle \otimes |j_r\rangle \otimes |j_\theta\rangle
    \label{eq:mapping}
\end{equation}
where the computational basis state $|j_i\rangle$ corresponds to the binary representation of the spatial grid index $j_i$, effectively encoding the positional information into the qubits.

The time evolution can also be encoded in the above basis functions using the second order split-operator method, or the symmetrized Trotter-Suzuki decomposition method \cite{zalka_simulating_1998} 
in the context of quantum computing. 
The matrix representations of the potential and kinetic energy operators are diagonal in the DVR and FBR representations, respectively, so that they can be efficiently implemented by diagonal phase gates in the quantum circuit. 
The potential energy propagator $e^{-i\hat{V}\Delta t/2}$ spans across all system registers, 
and the kinetic energy propagators $e^{-i\hat{T}_{R}\Delta t}$, $e^{-i\hat{T}_{r}\Delta t}$, $e^{-i\hat{T}^R_{\mathrm{rot}}\Delta t}$, and $e^{-i\hat{T}^r_{\mathrm{rot}}\Delta t}$ span only parts of the registers. 
The wavefunction is transformed independently between DVR and FBR at the same time by the standard Quantum Fourier Transform (QFT) and the $U_\theta$ transformation matrix for the radial and angular coordinates, respectively.

\subsection{Absorbing potential and the dilation method}
The wavefunction in the dissociation coordinate $R$ is unbound, but the size of the basis functions or the coordinate grids is 
finite. In order to enforce the outgoing boundary condition at the grid edge, a complex absorbing potential (CAP) is used to 
absorb the wavefunction before it reaches the grid boundary. 
Specifically, the CAP is often defined in the following polynomial form,
\begin{equation}
  V_{abs} = -i\alpha \left(\frac{R-R_{\mathrm{abs}}}{L_{\mathrm{abs}}}\right)^n
  \label{eq:abs}
\end{equation}
where $\alpha$ is the absorption strength, 
$R_{\mathrm{abs}}$ denotes the onset of the CAP, 
$L_{\mathrm{abs}}$ is its total length,
and $n$ is the degree of the polynomial.
This treatment effectively 
eliminates boundary-induced errors and allows the simulation to be performed 
within a truncated spatial domain. By confining the wavefunction to a limited region, 
the number of required grid points is significantly 
reduced, thereby minimizing the computational time and memory. 

Note that the CAP is a purely imaginary function, and the resulting propagator is a decaying
function, i.e., 
\begin{equation}
\hat{M}=e^{-iV_{\mathrm{abs}}\Delta t}=e^{-\alpha \left(\frac{R-R_{\mathrm{abs}}}{L_{\mathrm{abs}}}\right)^n \Delta t},
\end{equation}
which is not a unitary operator. 
To apply the CAP to the wavefunction is straightforward on classical computers. It is diagonal in the grid representation and can be applied to the wavefunction in the same way as the potential energy operator.

\begin{figure}[H]
  \centering
\begin{quantikz}[row sep=0.3cm, column sep=0.4cm]
\lstick{$\text{q}_r$}      & \gate[3]{e^{-iV\Delta t/2}} & \qw             & \gate[2]{e^{-iT_{\text{tot}}^r\Delta t}} & \qw                                      & \gate{\text{QFT}} & \gate{e^{-iT_r\Delta t}} & \gate{\text{QFT}^\dagger} & \gate[3]{e^{-iV\Delta t/2}} & \qw                             & \qw \\
\lstick{$\text{q}_\theta$} &                             & \gate{U_\theta} &                                          & \gate[2]{e^{-iT_{\text{tot}}^R\Delta t}} & \qw               & \qw                      & \gate{U_\theta^\dagger}   &                             & \qw                             & \qw \\
\lstick{$\text{q}_R$}      &                             & \qw             & \qw                                      &                                          & \gate{\text{QFT}} & \gate{e^{-iT_R\Delta t}} & \gate{\text{QFT}^\dagger} &                             & \gate[2]{\text{U}_{\text{dil}}} & \qw \\
\lstick{$\ket{0}_a$}       & \qw                         & \qw             & \qw                                      & \qw                                      & \qw               & \qw                      & \qw                       & \qw                         &                                 & \meter{}
\end{quantikz}
  \caption{Quantum circuit for a single propagation step $\Delta t$. The 
  circuit implements the second-order split-operator method, utilizing the 
  Quantum Fourier Transform (QFT) and Legendre-DVR transformation 
  ($U_\theta$) to switch between the DVR and FBR. The non-unitary complex 
  absorbing potential is implemented via the dilation unitary $U_\mathrm{dil}$, 
  which requires an additional ancilla qubit $|0\rangle_a$.}
  \label{fig:3D_qc}
\end{figure}
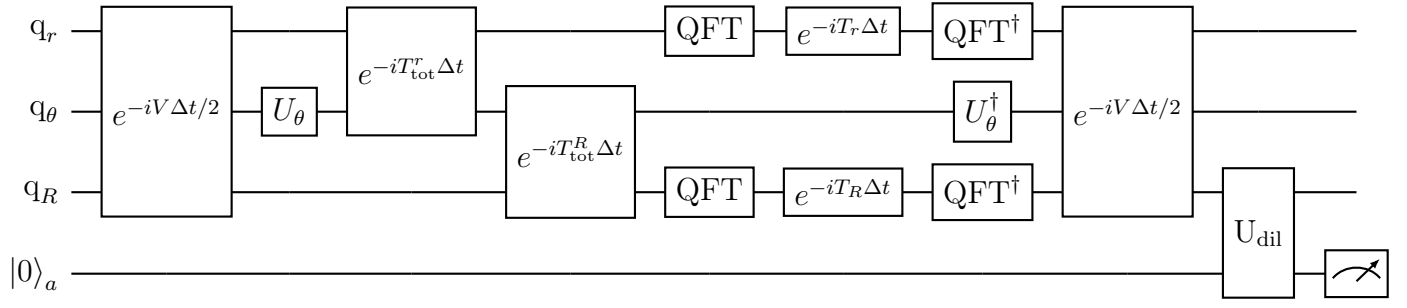

To apply the $\hat{M}$ operator on quantum computers is not straightforward, because the operations on quantum computers must be unitary.
To implement this non-unitary operator on a standard unitary
quantum circuit architecture, the dilation
scheme \cite{mangin-brinet_efficient_2024} can be used to
introduce an additional dilation ancilla qubit
$|0\rangle_a$ so that the non-unitary operator $\hat{M}$ is
embedded into an expanded unitary matrix $U_\mathrm{dil}$,
defined on the joint system-ancilla space as follows:
\begin{equation}
  U_\mathrm{dil} = \begin{pmatrix} 
  M & \sqrt{I - M^\dagger M} \\ 
  \sqrt{I - M^\dagger M} & -M 
  \end{pmatrix}
  \label{eq:dilation_unitary}
\end{equation}
Applying $U_\mathrm{dil}$ to the expanded space yields:
\begin{equation}
  U_\mathrm{dil}(|0\rangle_a \otimes |\Psi\rangle) = 
  |0\rangle_a \otimes (M|\Psi\rangle) + |1\rangle_a 
  \otimes (\sqrt{I - M^\dagger M}|\Psi\rangle)
  \label{eq:dilation_result}
\end{equation}
Measuring the dilation ancilla qubit at each step
allows for selecting of the wavefunction where the CAP is successfully appiled.
Specifically, a
measurement result of $|0\rangle_a$ 
projects the system onto the state
$\hat{M}|\Psi\rangle$, and the probability
corresponds to the squared modulus of the absorbed wavefunction
($p_0 = |\hat{M}|\Psi\rangle|^2$). The full quantum
circuit layout implementing a single integrated step
$\Delta t$ of the split-operator method with the
non-unitary dilation component within the IBM Qiskit
framework \cite{qiskit2024} is summarized in
Fig. \ref{fig:3D_qc}.

\subsection{Calculation of photodissociation cross section} 
The primary physical observable to extract from this
process is the total photodissociation cross section
$\sigma^{tot}(E)$, which is evaluated by taking the
Fourier transform of the time-domain auto-correlation
function $C(t)=\langle\Psi(0)|\Psi(t)\rangle$:
\begin{equation}
  \sigma^{tot}(E)=\frac{\pi\nu}{3c\epsilon_0}
  2\text{Re}\left\{\int^\infty_0 e^{iEt}C(t)dt\right\}
  \label{eq:pcs}
\end{equation}
where $\nu$ is the incident photon frequency, $c$ is
the speed of light, $\epsilon_0$ is the vacuum
permittivity, and $E$ is the total energy. 

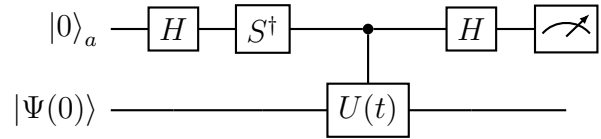
\begin{figure}[H]
  \centering
\begin{quantikz}[row sep=0.4cm, column sep=0.5cm]
\lstick{$\ket{0}_a$}     & \gate{H} & \gate{S^\dagger} & \ctrl{1}    & \gate{H} & \meter{} \\
\lstick{$\ket{\Psi(0)}$} & \qw      & \qw              & \gate{U(t)} & \qw      & \qw
\end{quantikz}
  \caption{Quantum circuit for the Hadamard test used to extract the 
  auto-correlation function. A control ancilla qubit $|0\rangle_a$ governs 
  the application of the total time evolution operator $U(t)$. }
  \label{fig:controlled_U}
\end{figure}

It is easy to calculate the auto-correlation function $C(t)$ on classical computers as the overlap between the initial wave function and the propagated wave function. The calculation on quantum computers is efficiently implemented by the Hadamard test, which is illustrated in Fig. \ref{fig:controlled_U}.
This procedure requires an additional ancilla qubit to
act as a control, which is initialized in $|0\rangle_a$
and transformed into the superposition
$(|0\rangle_a + |1\rangle_a)/\sqrt{2}$ via a
Hadamard gate. 
A controlled full-time
propagator $\mathcal{C}\text{-}U(t)$ is then applied
to the system, where $U(t) = [\hat{U}(\Delta
t)]^{N_s}$ represents the accumulated sequence of
$N_s$ propagation steps ($t = N_s \Delta t$).
Following a second Hadamard gate on the control ancilla,
the probability $P(0)$ of measuring it in the
$|0\rangle_a$ state yields the real component of
the overlap,
\begin{equation}
    P(0) = \frac{1}{2} + \frac{1}{2} \text{Re}
    \langle\Psi(0)|U(t)|\Psi(0)\rangle
\end{equation}
so that $\text{Re}\{C(t)\} =
2P(0) - 1$. The imaginary part
$\text{Im}\{C(t)\}$ is extracted in an identical
fashion by inserting an $S^\dagger$ phase gate after
the first Hadamard gate to rotate the complex
component into the measurement axis. 
It should be noted that the propagation of the wavefunction includes the non-unitary operator $\hat{M}$,
so that the measurements of the Hadamard test are post-selected on the conditions of obtaining a
successful $|0\rangle$ measurement from the dilation ancilla qubit.

\section{Results and Discussion}
\begin{figure}[H]
    \centering
    \includegraphics[width=0.8\textwidth]{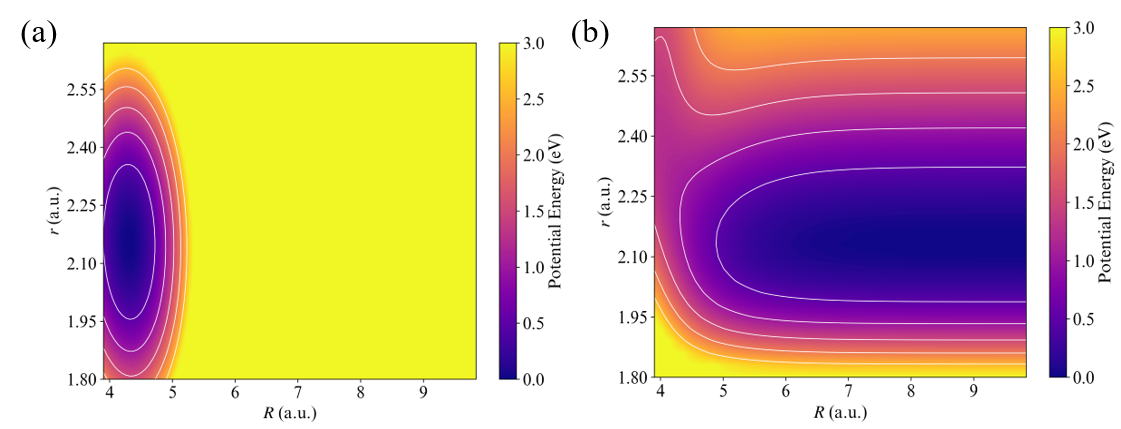}
    \caption{Two-dimensional contour plots of the potential energy 
    surfaces (PESs) for NOCl as a function of the radial 
    coordinates $R$ and $r$, with the Jacobi angle fixed at its 
    equilibrium value of $\theta = 127^\circ$. (a) The ground 
    electronic state ($S_0$) PES. (b) The first excited electronic state 
    ($S_1$) PES. Energy values are given in electron volts (eV).}
    \label{fig:pes}
\end{figure}

In this section, we present our calculated results of the 
photodissociation process of NOCl. 
The PES for both the $S_0$ and the $S_1$ electronic states are depicted in 
Fig. \ref{fig:pes}. For visualization purposes, the angle $\theta$ is fixed 
at $127^\circ$, which corresponds to the equilibrium geometry of the NOCl 
molecule. 
On the one hand, the ground state PES \cite{mcdonald_raman_1986} is a bound 
potential used to determine the initial wavefunction,
which in the present work was calculated on the classical computer by the Lanczos \cite{lanczos_iteration_1950}
algorithm and could also be calculated on quantum computers by the variation quantum algorithm. Here, we focus on the real-time evolution of the quantum system. 
On the other hand, the excited state PES \cite{manthe_wave-packet_1992} is purely repulsive 
along the $R$ coordinate, driving the rapid dissociation upon photon absorption.

% Start and end
The traditional TDWP method and the quantum algorithm are used to simulate the photodissociation of NOCl with the same parameters. The dissociation coordinate $R$ ranges from 
3.9 to 9.9 Bohr, and is discretized by $N_R = 128$ Fourier grid points, which are encoded into 
a dedicated quantum register comprising $n_R = 7$ qubits. The vibrational coordinate $r$ ranges from 1.8 to 2.7 Bohr, utilizing $N_r = 8$ grid points mapped onto $n_r = 3$ qubits. The angular degree of freedom $\theta$ is represented by $n_\theta = 5$ qubits, 
corresponding to $N_\theta = 32$ Legendre-DVR points. To enforce the 
outgoing boundary condition, the complex absorbing potential is positioned 
at $R_{\mathrm{abs}} = 8.0$ Bohr with an absorption strength of $\alpha = 0.2$ 
and a polynomial degree of $n = 2$. The real-time wave packet propagation 
is executed with a time step of $\Delta t = 2.0$ up to a total evolution 
time of $T_{\mathrm{tot}} = 1200.0$. The underlying quantum circuits are 
simulated utilizing the "density\_matrix" methods within the Qiskit Aer framework.

The performance of the proposed quantum computing algorithm in the Theory section is 
evaluated by analyzing two key observables: the auto-correlation function 
$C(t)$ and the resulting photodissociation cross section $\sigma^{tot}(E)$. 
This section is organized as follows: first, we validate our quantum 
algorithm on the 3D NOCl system by comparing the results implemented with IBM's Qiskit \cite{qiskit2024}
simulator with traditional TDWP benchmarks. The effect of the numbers of sampling shots is compared. Subsequently, we reduce the system to a 2D model in order to evaluate the robustness of the algorithm 
under realistic hardware noise conditions.

\subsection{Noiseless simulations and statistical sampling error} 
\begin{figure}[H]
  \centering
  \includegraphics[width=0.7\textwidth]{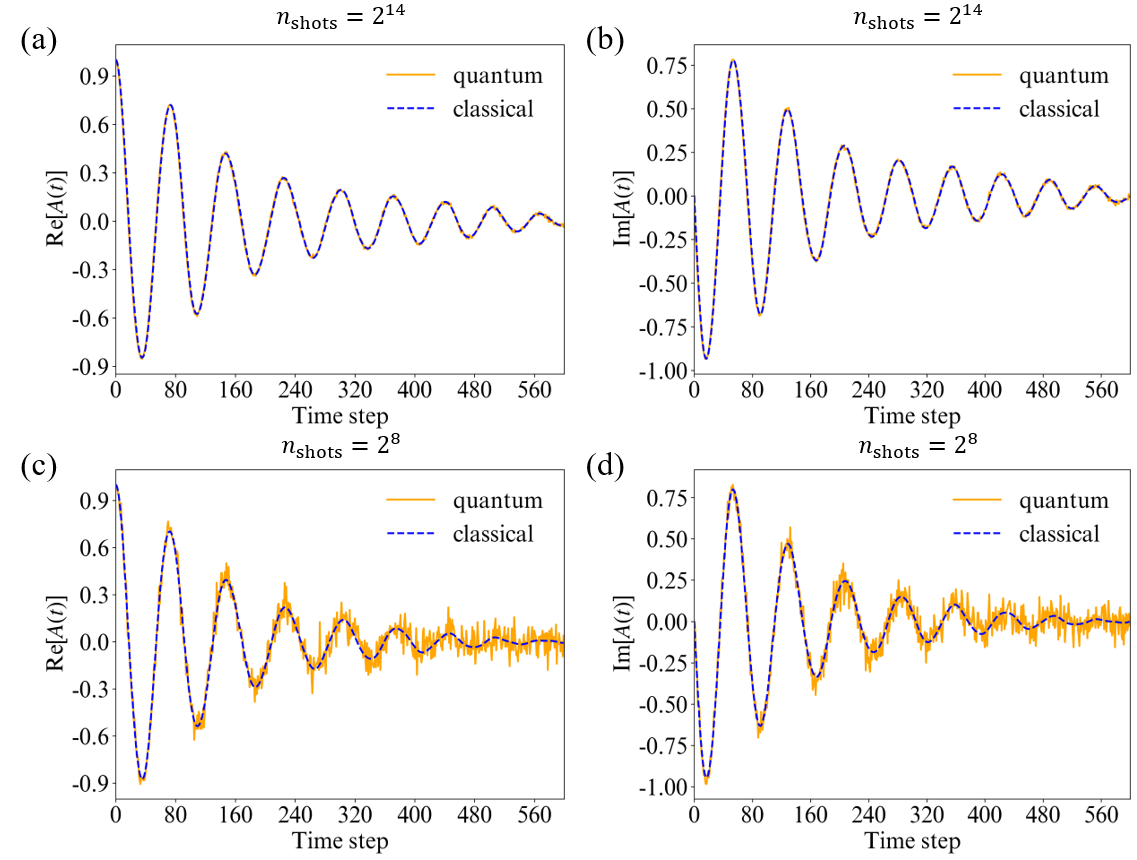}
  \caption{Comparison of the auto-correlation function $C(t)$ for the 3D NOCl 
  photodissociation process between quantum simulations (orange solid lines) 
  and classical TDWP benchmarks (blue dashed lines). Panels (a) and (b) 
  display the real and imaginary parts obtained from an ideal quantum 
  simulation with $2^{14}$ shots. Panels (c) and (d) display the 
  corresponding results obtained with a reduced sampling size of $2^8$ 
  shots, illustrating the impact of statistical fluctuations.}
  \label{fig:3d_acf}
\end{figure}

To verify the accuracy of the quantum algorithm described in the Theory 
section, we first performed calculations on the 3D NOCl system using the Qiskit 
Aer simulator with a large sampling number of $2^{14}$ shots. This setup 
minimizes statistical fluctuations, establishing a reliable baseline for 
comparison with classical TDWP benchmarks. The real and imaginary parts of 
the auto-correlation function $C(t) = \langle\Psi(0)|\Psi(t)\rangle$ were 
extracted using the Hadamard test.
As shown in Fig. \ref{fig:3d_acf}(a) and (b), the 
results (orange solid lines) with sufficient sampling shots exhibit excellent agreement with the classical 
TDWP benchmark results (blue dashed lines) over the entire time range. The rapid 
decay of $|C(t)|$ is accurately captured, reflecting the fast dissociation of 
the NOCl molecule on the repulsive $S_1$ potential energy surface. The 
precise reproduction of the oscillatory behavior of $C(t)$ indicates the good performance of the dilation scheme for the non-unitary $\hat{M}$ operator. 
To investigate the sensitivity of the algorithm to 
statistical fluctuations, we then reduced the sampling number to $2^8$ shots. 
As illustrated in Fig. \ref{fig:3d_acf}(c) and (d), the auto-correlation 
function obtained with such limited shot counts exhibits significant high-frequency 
oscillations and deviations from the TDWP benchmark results. This noisy pattern is a 
direct consequence of the dominant sampling error inherent in quantum 
measurements with a limited number of measurement shots.

\begin{figure}[H]
  \centering
  \includegraphics[width=0.7\textwidth]{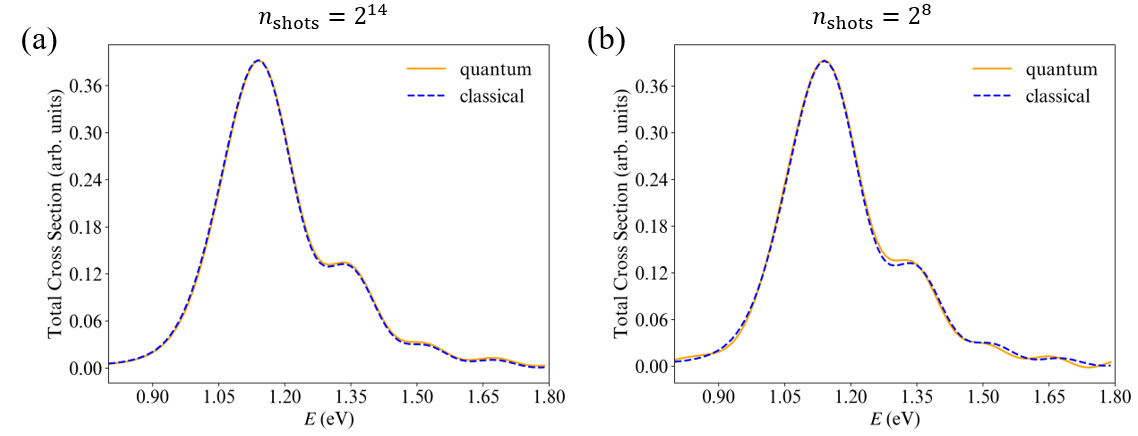}
  \caption{Comparison of the calculated total photodissociation cross 
  sections $\sigma^{\text{tot}}(E)$ for the 3D NOCl model. (a) Spectrum 
  derived from the ideal $2^{14}$ shots simulation. (b) Spectrum derived 
  from the $2^8$ shots simulation. Despite the significant sampling error in 
  the time domain, the essential spectral features are well-preserved after 
  the Fourier transform.}
  \label{fig:3d_pcs}
\end{figure}

The total photodissociation cross section $\sigma^{\mathrm{tot}}(E)$ is 
subsequently obtained by taking the Fourier transform of the calculated 
$C(t)$. Fig. \ref{fig:3d_pcs}(a) displays the resulting absorption spectrum 
for the ideal $2^{14}$ shots case. The quantum computing algorithm successfully 
reproduces the characteristic broad peak and subtle structures of the 
dissociation process, with peak positions and relative intensities closely 
matching the TDWP benchmark result.

Interestingly, with only $2^8$ shots, despite the distinct spikes in the auto-correlation 
function  (as seen in Fig. \ref{fig:3d_acf}c,d), 
the corresponding photodissociation cross section shown in 
Fig. \ref{fig:3d_pcs}(b) remains in remarkably good agreement with the 
benchmark result. While the cross section displays minor deviations at the high-energy tail, 
the overall envelope and critical peak positions are well-preserved. 
The Fourier transform in Eq. (\ref{eq:pcs}) involves an integral over the 
entire time range and effectively averages out the high-frequency statistical 
fluctuations. This intrinsic resilience to sampling errors allows for the 
extraction of meaningful spectral features, even with highly constrained sampling shots.

\subsection{Hardware Noise Simulation}

To evaluate the feasibility of implementing our algorithm on near-term noisy 
devices, it is crucial to investigate its performance under realistic 
hardware noise conditions. For this purpose, we utilize the GuadalupeV2 
noise model from the Qiskit Aer library, which mimics the hardware 
characteristics of the IBM Quantum Guadalupe device, incorporating specific 
gate errors, readout errors, and thermal relaxation effects ($T_1, T_2$).

To reduce the computational time for the complex noise simulation, we consider a reduced two-dimensional (2D) model of NOCl by fixing $\theta = 127^\circ$. For these 2D calculations, we select a sampling size of $2^{10}$ shots. This specific number of shots is chosen because the maximum absolute error of the resulting photodissociation cross section is within 0.005 compared to the TDWP benchmark, ensuring that the accuracy is sufficiently good while keeping the simulation efficient. We apply the GuadalupeV2 noise model to the quantum circuit illustrated in Fig. \ref{fig:controlled_U}, which is responsible for extracting the auto-correlation function under realistic hardware constraints. 

\begin{figure}[H]
  \centering
  \includegraphics[width=0.7\textwidth]{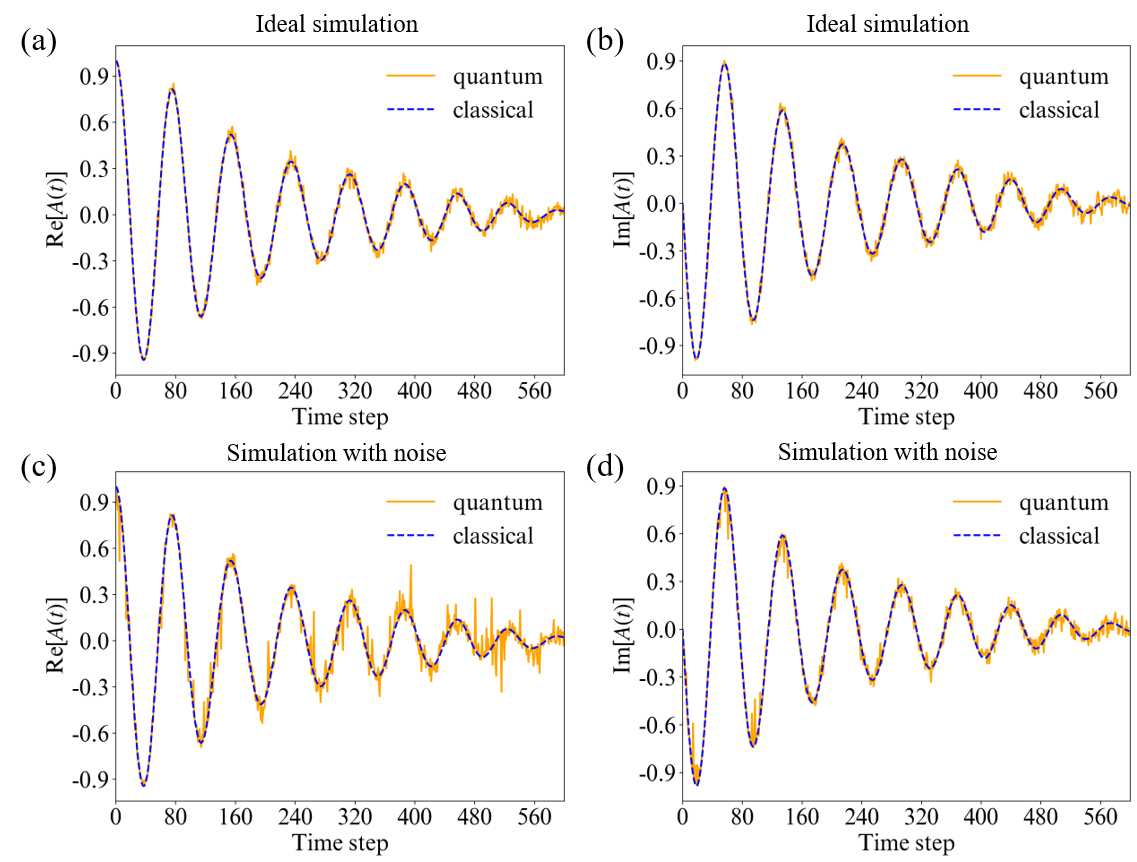}
  \caption{Auto-correlation function comparison for the 2D NOCl photodissociation 
  model. Panels (a) and (b) show the real and imaginary parts under an ideal 
  simulation with $2^{10}$ shots. Panels (c) and (d) display the results under 
  the realistic GuadalupeV2 noise model (also at $2^{10}$ shots), highlighting 
  the severe signal degradation caused by gate errors and decoherence.}
  \label{fig:2d_acf}
\end{figure}

As shown in Fig. \ref{fig:2d_acf}(a) and (b), the real and imaginary parts of the auto-correlation function obtained from the ideal quantum simulation with $2^{10}$ shots match the classical TDWP benchmarks well. In contrast to this ideal case, Fig. \ref{fig:2d_acf}(c) and (d) present the resulting noisy $C(t)$ under the GuadalupeV2 noise model. The noisy $C(t)$ exhibits severe amplitude degradation and significant fluctuations across the entire time evolution. These disturbances arise from the cumulative effect of quantum decoherence and imperfect gate operations inherent in the GuadalupeV2 device model, superimposed on the baseline statistical error.

\begin{figure}[H]
  \centering
  \includegraphics[width=0.7\textwidth]{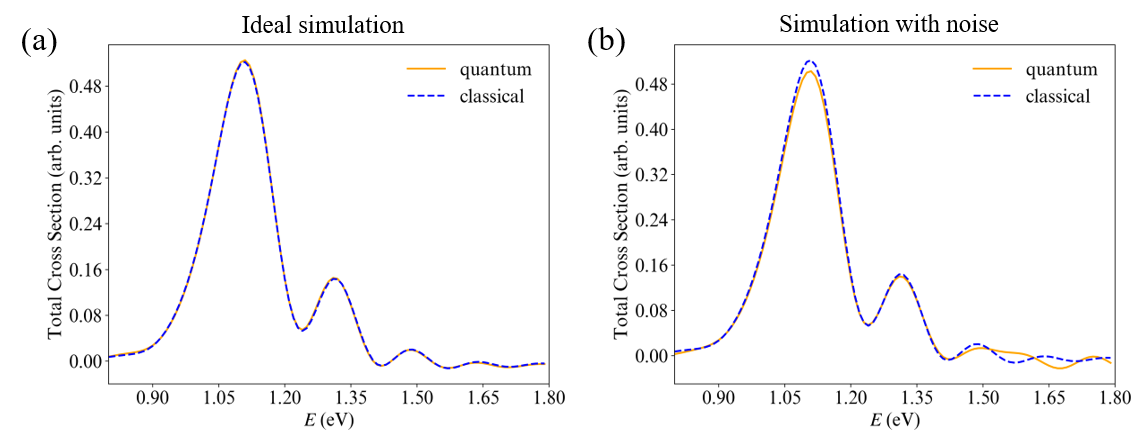}
  \caption{Photodissociation cross section comparison for the 2D NOCl model. 
  (a) Spectrum from the ideal $2^{10}$ shots simulation. (b) Spectrum from the 
  GuadalupeV2 noise model simulation. The Fourier transform effectively smooths 
  the hardware-induced noise, preserving the core physically meaningful features.}
  \label{fig:2d_pcs}
\end{figure}

Despite the marked degradation of the auto-correlation function in the time 
domain, the extracted photodissociation cross sections, shown in 
Fig. \ref{fig:2d_pcs}, still obtained meaningful results. The ideal 2D quantum 
simulation (Fig. \ref{fig:2d_pcs}(a)) accurately reproduces the peak 
structures and overall spectral envelope specific to the 2D NOCl model. 
Remarkably, the calculated cross section under the GuadalupeV2 noise model 
(Fig. \ref{fig:2d_pcs}(b)) remains consistent with the classical TDWP 
benchmark result. While the primary peak experiences a slight decrease in intensity 
and minor ripples appear in the high-energy tail, the central peak positions 
and the overall spectral profile are successfully obtained.

This result strongly highlights the intrinsic robustness of extracting 
frequency-domain observables from noisy time-domain quantum simulations. 
By integrating the degraded $C(t)$ over time, the high-frequency noise 
induced by near-term hardware imperfections is effectively filtered out. 
These observations provide compelling evidence that the proposed 
dilation-based TDWP algorithm is resilient to both sampling error 
and device noise, and suggest that the proposed algorithm can produce physically 
meaningful results on noisy devices.

\section{Conclusion}
In this work, we have developed and validated a quantum computing algorithm for 
the calculations of the photodissociation dynamics of the NOCl molecule. By 
integrating the split-operator propagation method with the dilation scheme, we 
successfully mapped the time evolution of the wavefunction in a finite coordinate range onto a 
quantum circuit with a finite number of qubits, and extracted the auto-correlation function through the 
Hadamard test. Our results demonstrate that the proposed algorithm accurately captures the
auto-correlation function of the NOCl system. Validation against classical 
TDWP benchmark results confirms the good performance of the algorithm when sufficient sampling shots are performed. Furthermore, our studies of the statistical and noise model simulation 
reveal a critical insight that insufficient sampling shots and device noise introduce significant fluctuations in the auto-correlation function $C(t)$, 
but the subsequent Fourier transform exerts a profound smoothing effect. 
This intrinsic resilience of the cross section allows for obtaining meaningful 
results even under the constraints of limited sampling shots
and realistic hardware noise (GuadalupeV2 model). 

These findings suggest that quantum computing is a promising approach for 
studying molecular photodissociation on the current noisy quantum devices. 
Even though scaling to larger systems remains challenging, 
especially the large circuit depths and gate counts required for encoding the high-dimensional potential energy operator, 
our results underscore the potential application of existing quantum hardware to quantum dynamics studies.
Furthermore, rapid advancements in both quantum hardware and algorithms will continue to provide viable pathways to overcome the exponential scaling inherent 
in the traditional time-dependent wavepacket calculation.

\begin{acknowledgement}
This work was supported by the National Natural Science Foundation of China (22288201, 22241301), the Strategic Priority Research Program of the
Chinese Academy of Sciences (XDB0970200), the Innovation Program for Quantum Science
and Technology (2021ZD0303305), the Guangdong Innovative and Entrepreneurial Research Team Program (2023ZT10Y058), the open fund (SKLMRD-K202623) of the state key
laboratory of chemical reaction dynamics in DICP, CAS. We gratefully acknowledge the
computing time supported by the Center for Computational Science and Engineering at the
Southern University of Science.

\end{acknowledgement}

\begin{suppinfo}

\end{suppinfo}

\bibliography{achemso-demo}

\end{document}